\documentclass[11pt]{article}
\usepackage{epsfig}
\usepackage{latexsym}
\usepackage{color}
\usepackage{amsmath,amssymb}
\usepackage{cite}
\usepackage{subfigure}
\makeatletter
\def\section{\@startsection {section}{1}{\z@}{-3.5ex plus -1ex minus
 -.2ex}{2.3ex plus .2ex}{\large\bf}}
\def\subsection{\@startsection{subsection}{2}{\z@}{-3.25ex plus -1ex
minus -.2ex}{1.5ex plus .2ex}{\normalsize\bf}}
\makeatother
\newcommand{\captionfonts}{\small}
\makeatletter  % Allow the use of @ in command names
\long\def\@makecaption#1#2{%
  \vskip\abovecaptionskip
  \sbox\@tempboxa{{\captionfonts #1: #2}}%
  \ifdim \wd\@tempboxa >\hsize
    {\captionfonts #1: #2\par}
  \else
    \hbox to\hsize{\hfil\box\@tempboxa\hfil}%
  \fi
  \vskip\belowcaptionskip}
\makeatother   % Cancel the effect of \makeatletter
\catcode`@=11
\def\marginnote#1{}
\newcount\hour
\newcount\minute
\newtoks\amorpm
\hour=\time\divide\hour
by60
\minute=\time{\multiply\hour by60 \global\advance\minute
by-\hour}
\edef\standardtime{{\ifnum\hour<12 \global\amorpm={am}
\else\global\amorpm={pm}\advance\hour by-12 \fi
 \ifnum\hour=0
\hour=12 \fi
 \number\hour:\ifnum\minute<10
0\fi\number\minute\the\amorpm}}
\edef\militarytime{\number\hour:\ifnum\minute<10
0\fi\number\minute}
\def\draftlabel#1{{\@bsphack\if@filesw
{\let\thepage\relax
 \xdef\@gtempa{\write\@auxout{\string
\newlabel{#1}{{\@currentlabel}{\thepage}}}}}\@gtempa
 \if@nobreak
\ifvmode\nobreak\fi\fi\fi\@esphack}
\gdef\@eqnlabel{#1}}
\def\@eqnlabel{}
\def\@vacuum{}
\def\draftmarginnote#1{\marginpar{\raggedright\scriptsize\tt#1}}
\def\draft{\oddsidemargin
0.0truein
 \def\@oddfoot{\sl preliminary draft \hfil
\rm\thepage\hfil\sl\today\quad\militarytime}
 \let\@evenfoot\@oddfoot
\overfullrule 3pt
 \let\label=\draftlabel
\let\marginnote=\draftmarginnote
\def\@eqnnum{(\theequation)\rlap{\kern\marginparsep\tt\@eqnlabel}
\global\let\@eqnlabel\@vacuum}
}
\catcode`@=12
\def\dj{\hbox{d\kern-0.347em \vrule width 0.3em height 1.252ex depth
-1.21ex \kern 0.051em}}

\def\ee{{\rm e}\,}

\def\Dirac{\,\raise.15ex\hbox{/}\mkern-13.5mu D}
\def\dirac{\,\raise.15ex\hbox{/}\kern-.57em \partial}
\def\aslash{\,\raise.15ex\hbox{/}\mkern-13.5mu A}
\def\shalf{{\ifinner {\textstyle \frac{1}{2}}\else \frac{1}{2} \fi}} 
\def\sthreehalfs{{\ifinner {\textstyle \frac{3}{2}}\else \frac{3}{2} \fi}} 
\def\sshalf{{\ifinner {\scriptstyle \frac{1}{2}}\else \frac{1}{2} \fi}} 
\def\sfourth{{\ifinner {\textstyle \frac{1}{4}}\else frac{1}{4} \fi}}
\def\sphifour{{\ifinner {\textstyle \frac{1}{4!}}\else \frac{1}{4!} \fi}}
\def\lsim{\stackrel{<}{_\sim}}

\def\XXint#1#2#3{{\setbox0=\hbox{$#1{#2#3}{\int}$}
     \vcenter{\hbox{$#2#3$}}\kern-.5\wd0}}

\def\bea{\begin{eqnarray}} \def\eea{\end{eqnarray}}
\def\be{\begin{eqnarray}} \def\ee{\end{eqnarray}} 
\newcommand{\promille}{%
  \relax\ifmmode\promillezeichen
        \else\leavevmode\(\mathsurround=0pt\promillezeichen\)\fi}
\newcommand{\promillezeichen}{%
  \kern-.05em%
  \raise.5ex\hbox{\the\scriptfont0 0}%
  \kern-.15em/\kern-.15em%
  \lower.25ex\hbox{\the\scriptfont0 00}}
\newcommand{\beq}{\begin{eqnarray}}
\newcommand{\eeq}{\end{eqnarray}}

\newcommand{\gsim}{\raisebox{-0.13cm}{~\shortstack{$>$ \\[-0.07cm]
      $\sim$}}~}

\newcommand{\comment}[1]{}
\begin{document}

\thispagestyle{empty}

\begin{flushright}
\flushright
ITP-UU-11/17\\
SPIN-11/11
\end{flushright}

\begin{center}

\vspace{1.7cm}

{\LARGE\bf 
Higgs Physics at the Large Hadron Collider\footnote{Written version of the 
talk presented at the 11th Workshop on High Energy Physics Phenomenology 
(WHEPP), Allahabad, India.}
}

\vspace{1.4cm}

{\bf Rohini M. Godbole}

\vspace{1.2cm}

Center for High Energy Physics, Indian Institute of Science,
Bangalore 560 012, India.\\
\&\\
Institute for Theoretical Physics and Spinoza Institute,
Utrecht University, 3508 TD Utrecht,\\
The Netherlands.
\end{center}

\centerline{\bf Abstract}
\vspace{2 mm}
\begin{quote}
\small
In this talk I will begin by summarising the importance of the Higgs physics
studies at the LHC. I will then give a short description of the pre-LHC 
constraints on the Higgs mass and the theoretical predictions for the LHC 
along with a discussion of the current experimental results, ending with 
prospects in the near future at the LHC. In addition to the material covered
in the presented talk, I  have included in the writeup, a critical appraisal 
of the theoretical uncertainties in the Higgs cross-sections at the Tevatron  
as well as a discussion of the recent experimental results from the LHC which 
have become available since the time of the workshop.
\end{quote}

\section{Introduction}
\begin{figure}[!hbtp]
\centering
\subfigure[]
{
\includegraphics*[width=6cm,height=3cm]{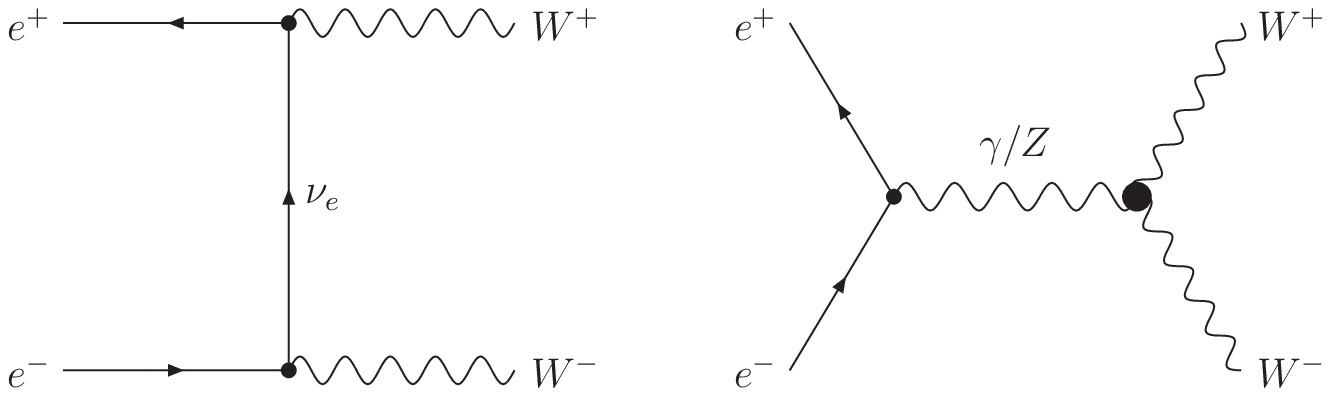}
\label{fig1a}
}
\subfigure[]
{
\includegraphics*[width=8cm,height=6cm]{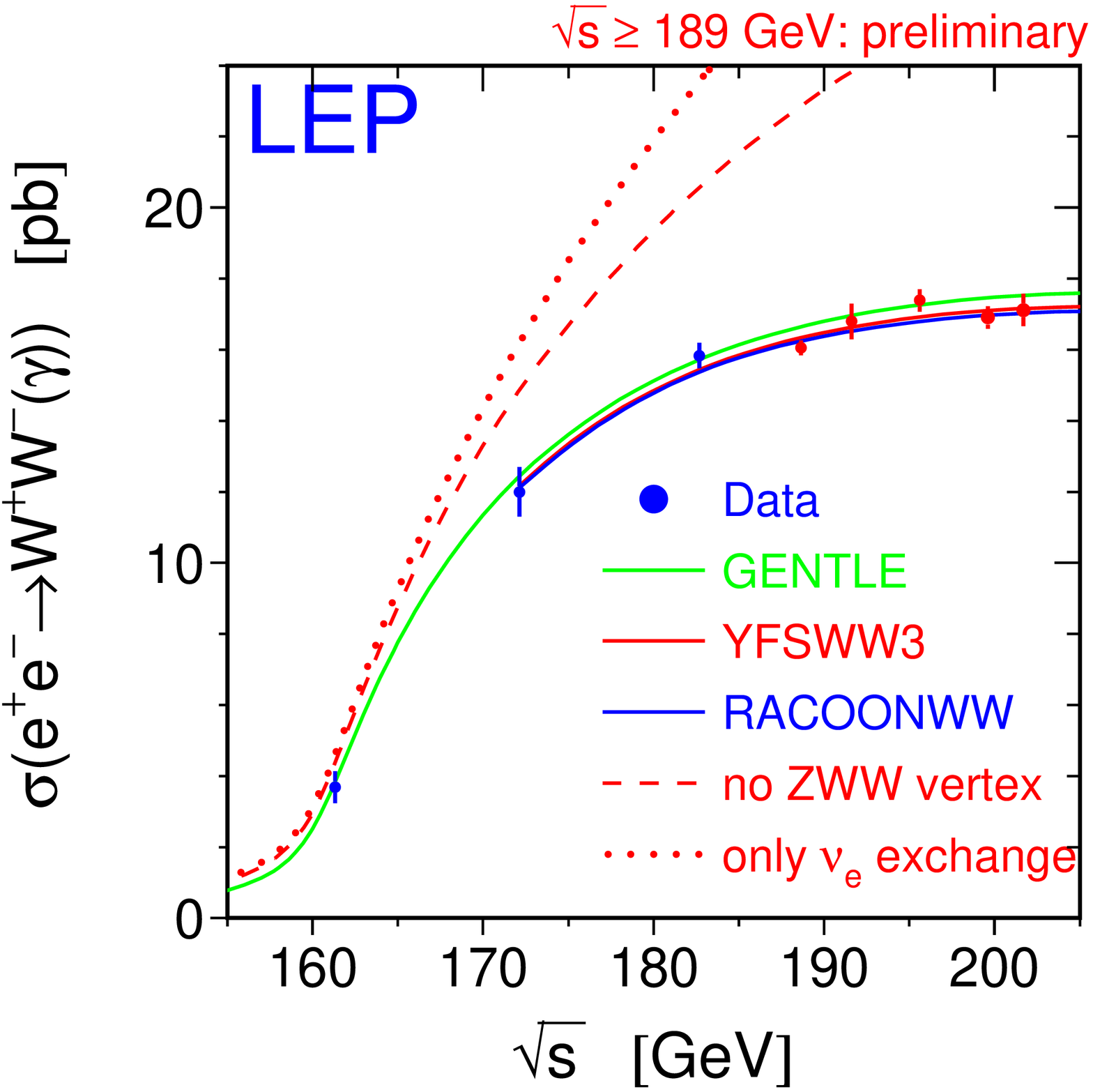}
\label{fig1b}
}
\caption{Comparison of the LEP data (taken from LEPEWWG)  with the SM 
prediction shown in (b), the contributory processes being depicted in  
(a).}
\label{fig1}
\end{figure}

It goes without saying that establishing the exact nature of the mechanism
of electroweak symmetry breaking is perhaps 'THE'  most important issue
in the subject of particle physics at present and arguably the 'raison 
d'\^etre'  for the Large Hadron Collider (LHC). The  excellent agreement of
the LEP data on $\sigma(e^+ e^- \rightarrow W^+ W^-) $  with predictions of 
the Standard Model (SM) shown in Fig.~\ref{fig1}
gives us a direct confirmation of the triple gauge boson ($ZWW$)
coupling  as predicted by  the $SU(2)\times U(1)$ symmetry. At the same time
the observed nonzero mass of the $W$--boson confirms that the same EW symmetry
is broken as well. The Higgs mechanism~\cite{Higgs,SM} is one way of achieving 
the desired breakdown of the EW symmetry.  This predicts the existence of a 
$J^{PC} = 0^{++}$ state, as the remnant of the $SU(2)_L$ doublet, with
{\it precise predictions} for the coupling of this state to all the
SM particles, but is able to give only very {\it weak theoretical constraints} 
on its mass. Since this is the only particle of the SM~\cite{SM} still lacking
confirmation by direct experimental observation, it is clear that discovery of 
the Higgs boson  and a study of its properties are at the heart of the LHC 
program which has begun operations at 7 TeV since February 2010. 

A few remarks are in order here. Theoretical ideas of electro weak symmetry 
breaking (EWSB) span a large range, beginning from the weakly coupled Higgs
to those of  strong interaction dynamics which can involve  a composite (or
worse no) Higgs boson.  All of these, including the SM, of course have had 
to pass the acid test of the EW precision measurements, at the Z 
pole at the LEP collider.  The latter 
class of models, involving  dynamical symmetry breaking triggered by strong
dynamics, have got a new lease of life due to theoretical developments in
the context of models with extra dimensions.  Needless also to say that the
contents of the Higgs sector and the properties of the said particles, both can
differ from the SM in the many different proposals of going beyond the standard 
model (BSM). Further, CP violation in the Higgs sector can be the possible BSM 
physics that seems to be necessary for a {\it quantitative} explanation of the 
Baryon Asymmetry in the Universe. In addition to this, the dark matter in the 
Universe, seems to also not consist of any of the known particles in the SM. 
Interestingly, almost all the extensions of the SM, always have a particle 
which has all the right properties to be a dark matter (DM) candidate. Since 
most of the extensions of the SM are introduced to deal with some of the, 
not yet completely understood and/or unsatisfactory, features of the  
EWSB, almost always this candidate DM particle has interesting connections to
Higgs physics as well.

We expect the LHC to unravel the secrets of the physics of the  EWSB as well as to provide pointers to the BSM physics  which, we hope, in turn will provide 
key 
to the explanation of the issues of cosmological importance, viz. the Baryon 
Asymmetry in the Universe (BAU) and the DM in the Universe. The discussion 
preceding these few lines, should then convince us that 'Higgs Physics at the 
LHC', will indeed touch upon almost all the aspects of active investigation in 
theoretical and experimental particle physics.

While discussing 'Higgs Physics at the LHC' the different issues that need 
be addressed are: 

\begin{enumerate}
\item Discovering the spin $0$ state(s) and measuring their mass 
as well as  couplings to fermions and gauge bosons.
\item Can these measurements uniquely decide the gauge group representation
to which these scalar(s) belong? Can they give information about whether the 
SM is a strongly coupled theory with (perhaps) a composite Higgs boson or a 
weakly coupled  theory with an elementary Higgs boson? 
\item Is there a CP violation in the Higgs sector?
\item Can the LHC give any information on the triple Higgs coupling, which is
present as a result of the spontaneous symmetry breaking? 
\end{enumerate}
LHC is capable of answering these questions to different degree of completeness,
some early and some in the far future.

Clearly, a short survey such as this can not do justice to the enormous amount 
of work done on the subject~\cite{higgs-review,lhc}. The discussion here will 
hence only focus on a few issues.  I shall summarise first the current 
constraints on the mass and then go on to discuss the status of theoretical
predictions for the LHC. I will then present the current projections for 
discovery and exclusions made by the two LHC experiments. I will then discuss 
briefly two new developments in the subject: 1) the jet substructure technique 
which enables  use  of the  $b \bar b$ final state arising from the 
Higgs decay. Due to the large QCD backgrounds this final state could not 
always be utilised in the analyses hitherto, 2) the  possibility of 
obtaining  spin and parity of the observed scalar state even in the early data.

\section{SM Higgs: profile and current constraints}
\begin{figure}[!hbtp]
\centering
\includegraphics*[width=10cm,height=6cm]{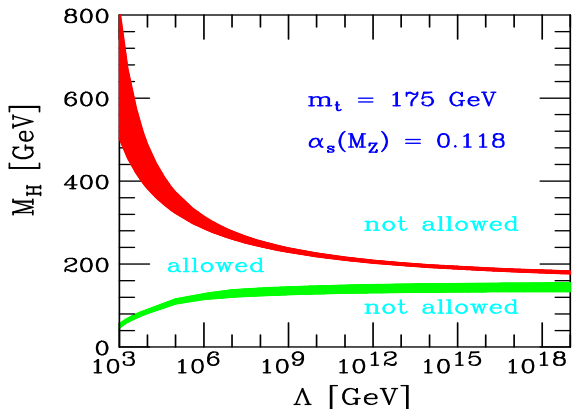}
\caption{
Theoretical upper and lower bounds on the Higgs mass in the SM from the 
assumption of the validity of the SM upto a cut--off scale $\Lambda$ and 
vacuum stability\cite{Hambye:1996wb}.}
\label{fig2}
\end{figure}

As is well known, theoretical considerations are capable of only giving bounds
on the Higgs mass. These bounds arise from considerations of triviality 
and vacuum stability of the Higgs sector and are shown in Fig.~\ref{fig2}.
To be specific, these limits are obtained from requiring that the Landau pole 
in the 
evolution of the Higgs self coupling lie above a cut off scale $\Lambda$, 
as well as by demanding that the scalar potential be bounded from below.
The limit on the Higgs mass of about 800 GeV for $\Lambda \sim 1$ TeV seen
in Fig.~\ref{fig2}, is also consistent with the bound that one  obtains on 
very general
grounds by demanding that the tree level scattering amplitudes involving the 
gauge bosons $W$ and $Z$ satisfy unitarity~\cite{H-LQT}. In fact it is
interesting to know that this unitarity demand had been used to derive 
successfully, the gauge and Higgs boson content of the SM and the coupling 
structure of these bosons among themselves and the fermions in the 
SM~\cite{unitarity-sm}.

\begin{figure}[!hbtp]
\centering
\subfigure[]
{
\includegraphics*[width=6cm,height=6.0cm]{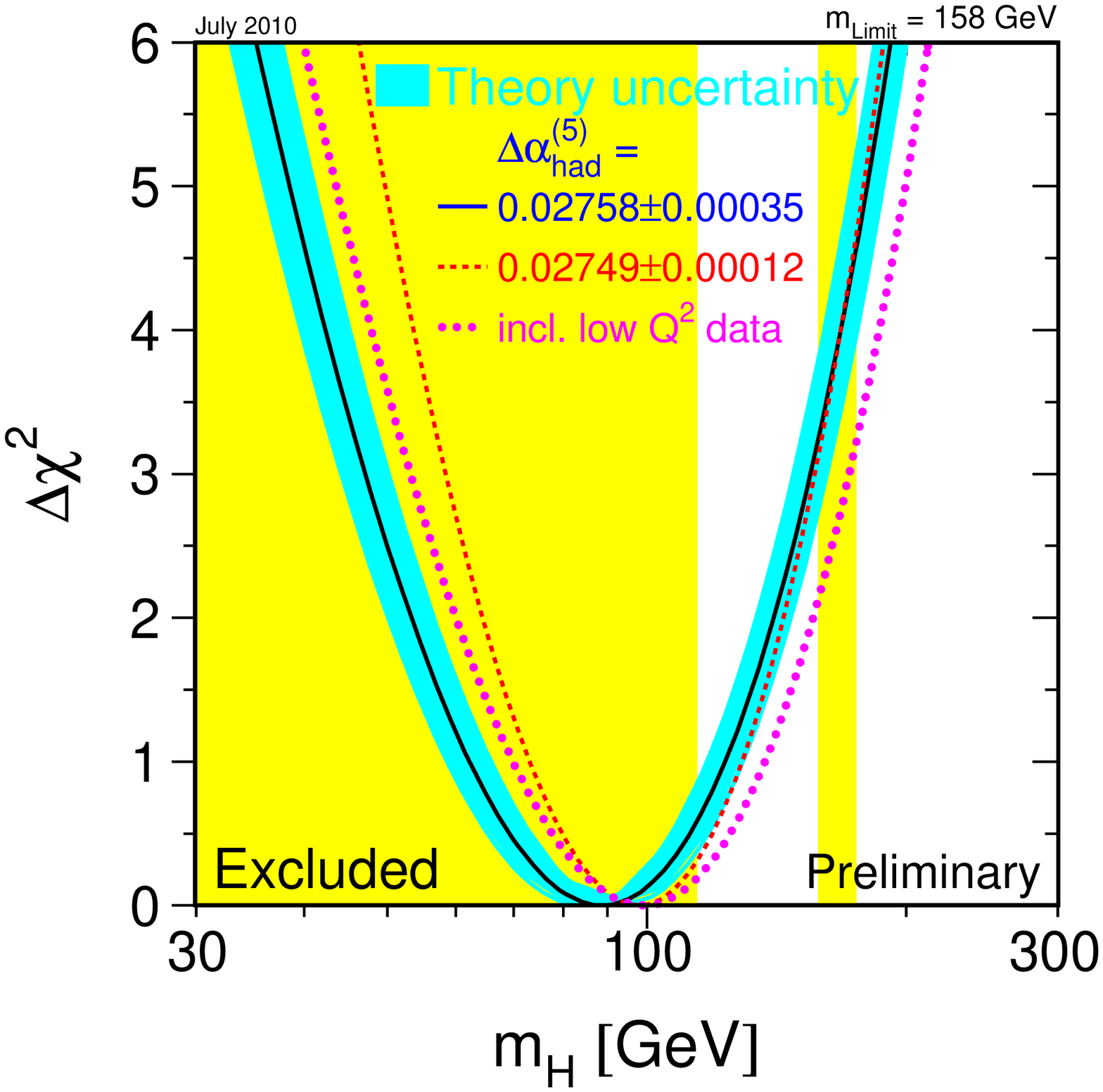}
\label{fig3a}
}
\subfigure[]
{
\includegraphics*[width=6cm,height=5.5cm]{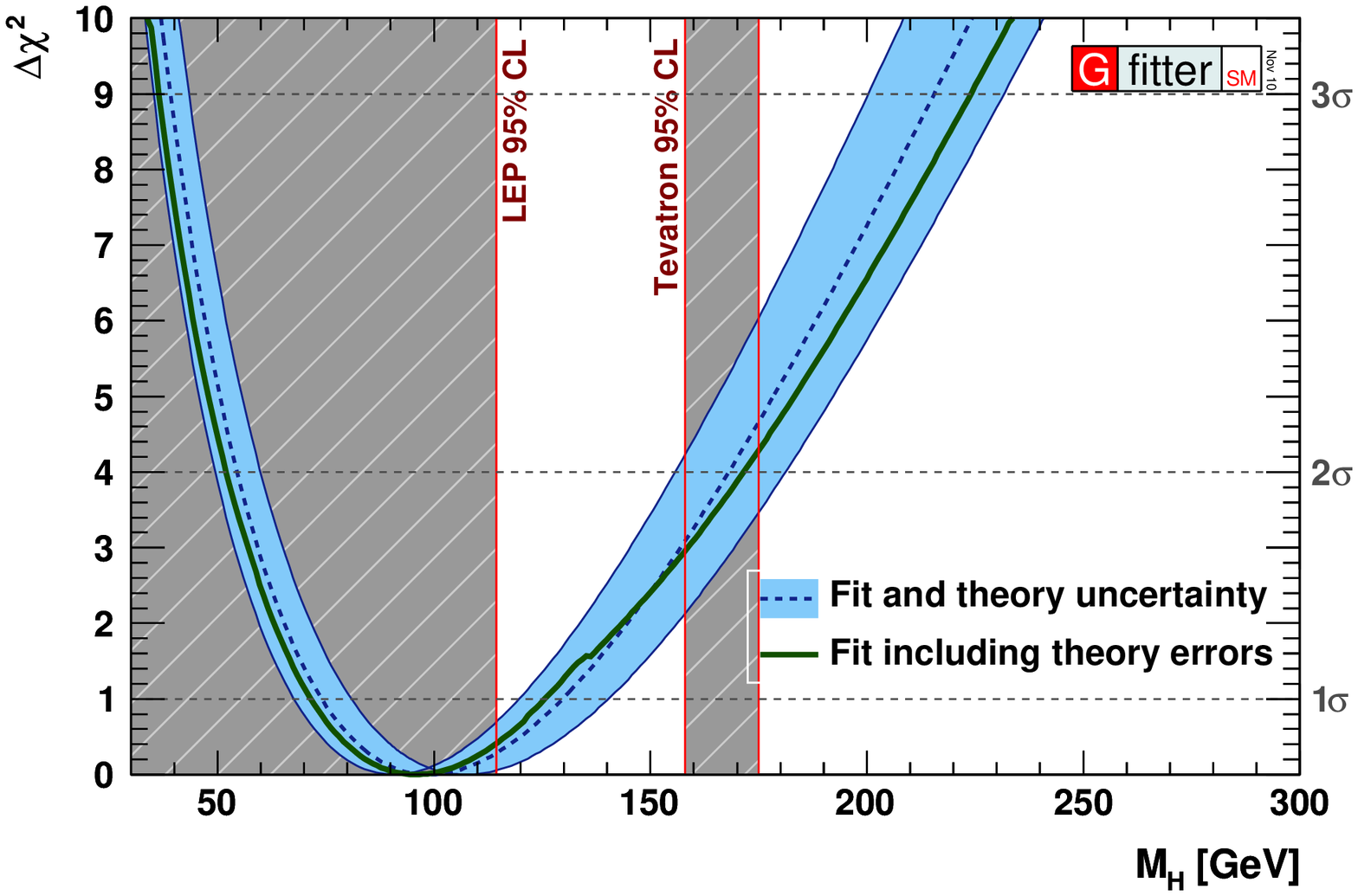}
\label{fig3b}
}
%\label{fig3}
\caption{The two panels show a summary of the current, direct and indirect, 
experimental constraints on the Higgs mass from the collider experiments, 
taken from the web pages of the LEPEWWG and the Gfitter group. Both
the panels show $\Delta \chi^2$ as a function of $M_H$ for a SM fit to a 
variety of precisely measured electroweak observables.}
\label{fig3}
\end{figure}

The radiative corrections to the $W/Z$ boson masses coming from the Higgs boson
are proportional to $\log (M_H/M_W)$. This logarithmic dependence is related to 
the 'custodial' symmetry which the SM possesses. As a result the precision 
measurements of the gauge boson masses already can put {\it indirect} 
constraints on the allowed value of  $M_H$. Further the 'direct' searches at 
LEP~\cite{leplimits}
and Tevatron~\cite{tevlimits} also exclude the existence of a Higgs boson, in 
certain mass regions. Fig.~\ref{fig3} shows a compilation of these indirect 
constraints obtained by the LEPEWWG and Gfitter groups, along with the 
limits from the direct searches. The two panels show $\Delta \chi^2$ 
as a function of $M_H$ for a SM fit to the various EW precision 
observables. Two comments are in order here. Firstly the quality of the
SM fit indicates that every  candidate for the BSM physics must have
at least an approximate custodial symmetry. Secondly, the exclusion of 
an SM Higgs in the mass range around $160$ GeV, coming from direct searches at
the Tevatron, has a nontrivial dependence on the nonperturbative knowledge of
the proton~\cite{us,moch}. These different issues can affect the estimate of
uncertainties in the theoretical predictions for the cross-sections for the 
signal as well as the background. These issues can thus affect the 
significance of this limit.  I will comment upon it later.

These results shown in Fig.~\ref{fig3} tell us that in the SM, current data 
prefer a light Higgs and on inclusion of  the direct limits from the collider 
searches, one gets $M_H < 185$ GeV, at $95 \%$ confindence level (CL). The 
closeness of this bound
with that seen in the  theoretical analysis presented in Fig.~\ref{fig2} 
in fact raises the hairy prospect that we might find only such a light Higgs 
with $M_H \leq 180$ --$185$ GeV  and nothing else at the LHC.  It should be 
mentioned here however, that some of 
the details of the analyses of the EW observables,  are quite sensitively 
dependent on the way in which the theoretical and experimental errors are 
accounted for therein. This knowledge thus sets now the stage for the LHC 
Higgs searches. 

A remark is in order before we proceed with a discussion of Higgs searches
at the LHC in the mass range allowed by the current direct and indirect 
constraints.  These indirect and direct bounds of Figs.~\ref{fig2} 
and \ref{fig3} indicate that just the mass of the observed scalar state, 
should be able to give information about the energy scale at which new physics 
may appear. For example, a  Higgs boson with $m_H \sim 130$--$140$ GeV, 
while not
providing any 'proof' for TeV scale physics, will indicate strong possibility
of TeV scale physics which can keep the Higgs 'naturally' light.
Supersymmetry is an example of one BSM physics which can achieve this.
Observation of a scalar state with mass in the 
region of $\sim 180$ GeV will  indicate possible compatibility with the 
absence of any new physics upto such high scales as may not be accessible 
in direct searches at the LHC. On the other hand a heavier Higgs 
(say $\sim  300$ GeV) would necessarily mean new physics around TeV scale. 
Specific examples of such new physics can be, e.g., a sequential fourth 
generation of fermions and/or presence of strong interaction dynamics 
around TeV scale. An interesting question to ask in the context of Higgs 
and BSM searches at the LHC is about the time line of possible discovery 
of a Higgs of a particular mass and that of the corresponding new 
physics which has been postulated to make the existence of a Higgs boson 
with that mass consistent with the different theoretically motivated 
requirements such as naturalness and/or better consistency with the EW 
precision measurements used in arriving at Fig.~\ref{fig3}.

\begin{figure}[!hbtp]
\centering
\subfigure[]
{\includegraphics*[width=6cm,height=5cm]{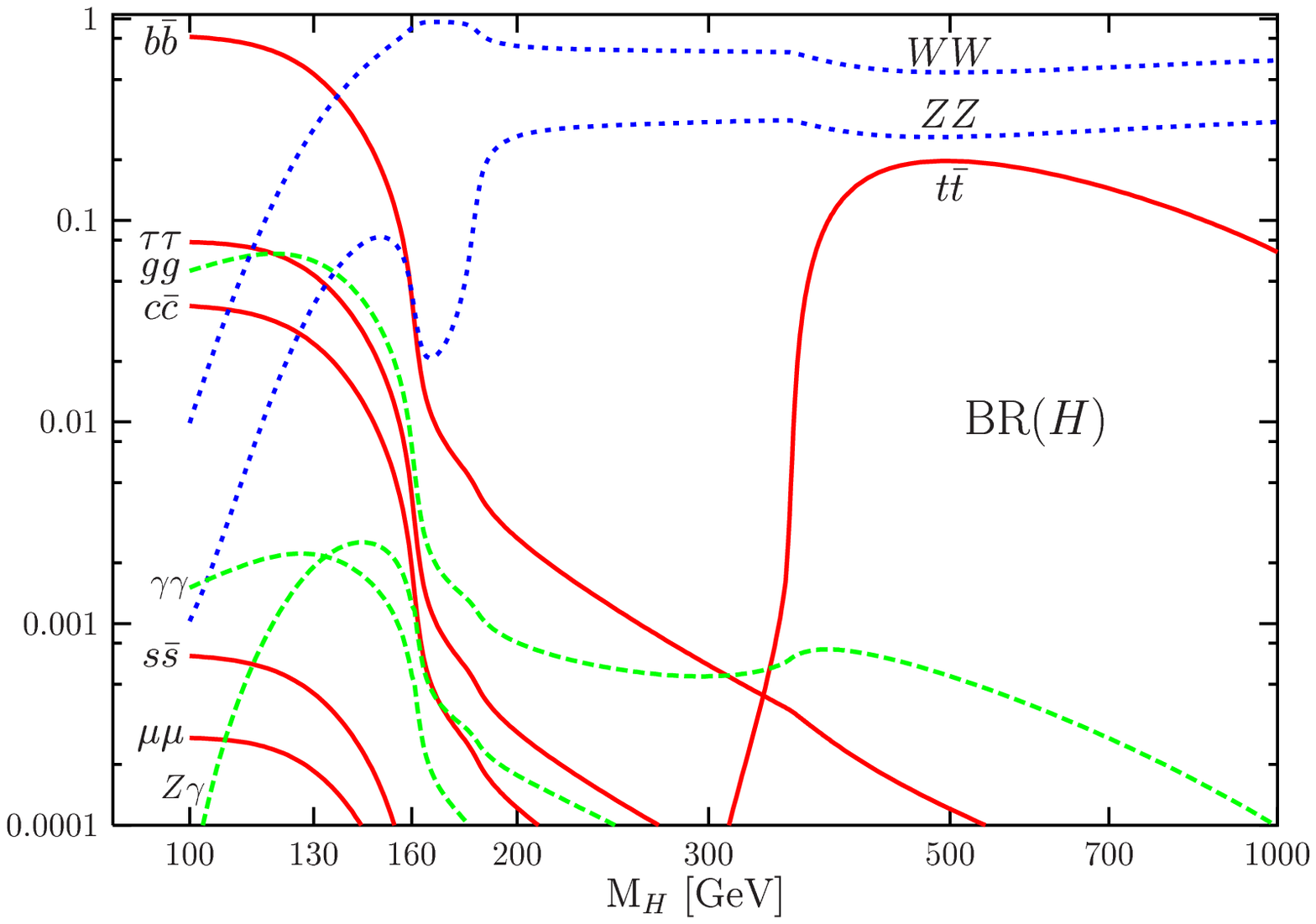}
\label{fig4a}}
\subfigure[]
{\includegraphics*[width=6cm,height=5cm]{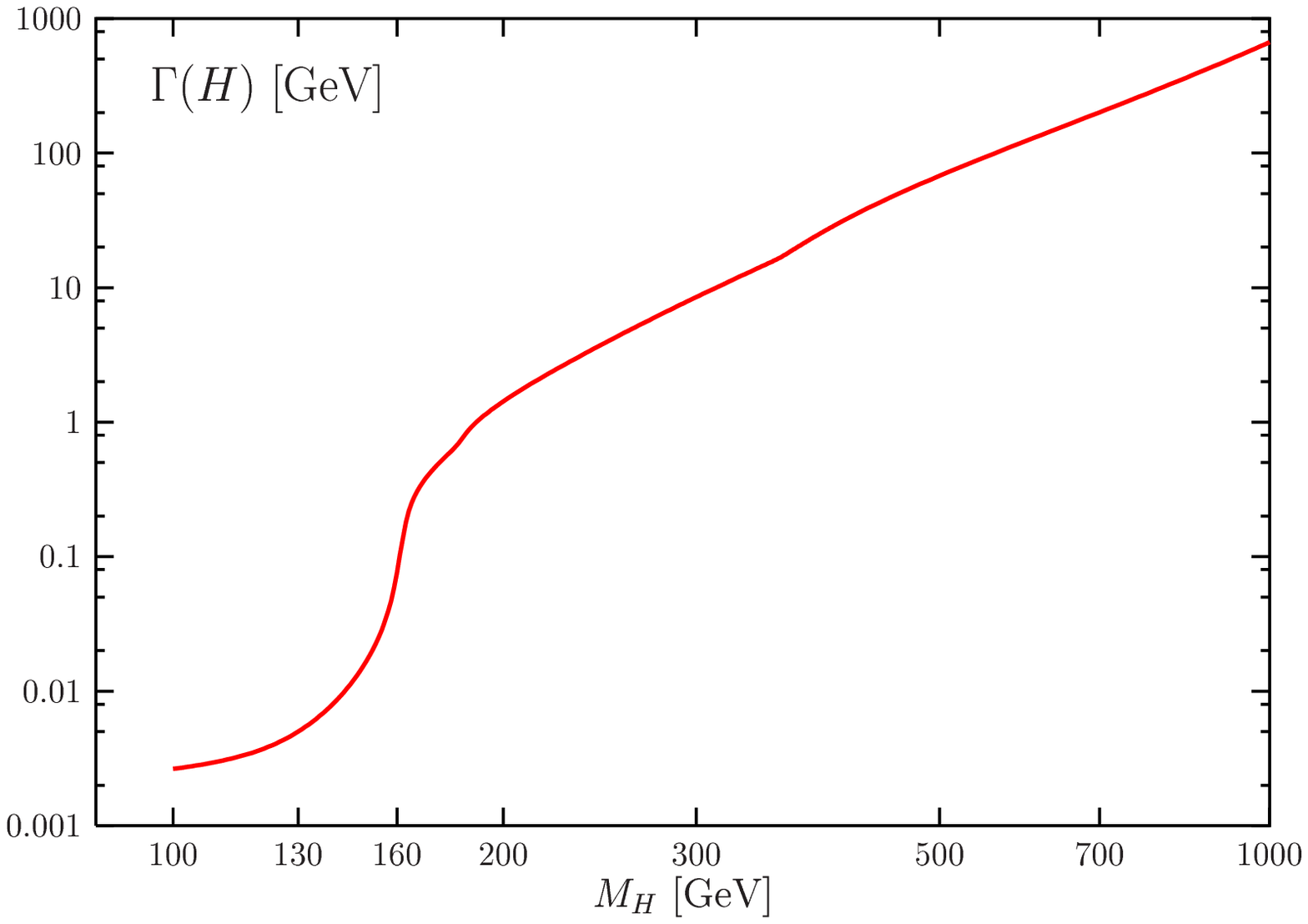}
\label{fig4b}}
\caption{The decay branching ratios and the total decay width 
of the SM Higgs boson as a function of its mass, taken 
from~\protect\cite{higgs-review}.}
\label{fig4}
\end{figure}

We begin by summarising some of the Higgs properties, relevant for its search 
at the LHC, such as its  width and branching ratio into different channels.  
Fig.~\ref{fig4} shows the branching ratios for the SM Higgs over the entire 
mass range that is consistent with the theoretical constants mentioned earlier.
Thus we see that for the light Higgs, such as the one indicated by above 
constraints,
the width of the Higgs boson is expected to be $\lsim 1$ GeV. For a  
Higgs with mass $\lsim 130$ GeV branching ratio into the $ b \bar b$ channel
is expected to be large, with that in the $\gamma \gamma$ channel  $\sim
10^{-3}$. For larger values of the Higgs mass, the $VV$ decay modes are 
dominant, with $WW$ and $ZZ$ sharing it in the ratio 2:1.  For the Higgs in 
the mass range $\gsim 135$ GeV the four fermion decay mode is the 
most important one. Combined QCD and EW corrections can change this by upto a 
few percent.  Due to the large QCD backgrounds, the $\gamma \gamma$  mode is 
considered optimal for the  Higgs in the mass range $115$--$135$ GeV. However, 
there has been a major change in the attitude since  it has 
been pointed out that the use of  $b \bar b$ final states can be made 
possible  using jet substructure methods~\cite{butterworth}. I will give a 
short description of these methods in the later discussions.

\section{Production of the Higgs at the LHC}
Since LHC is a hadronic collider, one of the most relevant activity is accurate
predictions of the expected cross-sections as well as differential 
distributions in 
important kinematical variables such as, eg.,  $p_T^H$  for various Higgs 
production processes. QCD factorisation theorem at short distances tells us
that this cross-section can be calculated in the following formalism:
\begin{eqnarray}
\sigma(pp \rightarrow X +..) &= \sum_{a,b}\int_0^1 dx_1 dx_2
               f_{a} (x_1,\mu_F^2) f_b(x_2,\mu_F^2)
               \nonumber \\
              & \times \sigma(a+b \rightarrow X)
                \left(x_1,x_2,  \mu_R^2,
               \alpha_s(\mu_R^2), \alpha(\mu_R^2),
            \frac{Q^2}{\mu_R^2},\frac{Q^2}{\mu_F^2}\right)
\end{eqnarray}
An accurate calculation requires precise inputs on two non perturbative 
quantities $\alpha_s$ and the parton density functions, PDF's, along with
an accurate evaluation of the subprocess cross-sections. An enormous amount of 
work has been done on the subject. An evaluation of the theoretical 
uncertainties in the predictions for the Higgs cross-section at the LHC was 
presented in ~\cite{Baglio:2010ae}.  In fact a joint collective  
effort~\cite{lhcxsection}, involving  experimentalists and theorists, has been 
made recently to make the most accurate predictions for total observable
cross-sections, taking into account most of the current theoretical 
uncertainties, 
both in the calculation of production cross-sections and the branching ratios. 
The next step is to do the same for exclusive distributions. Since all the
experimental searches necessarily require cuts on the phase space variables,
predictions for exclusive distributions for both the signal and dominant
background are extremely crucial.  Below, I summarise the main features and 
refer the reader to ~\cite{Baglio:2010ae,lhcxsection} and references therein 
for further details.

\begin{figure}[!hbtp]
\centering
\subfigure[]
{\includegraphics*[width=6cm,height=4cm]{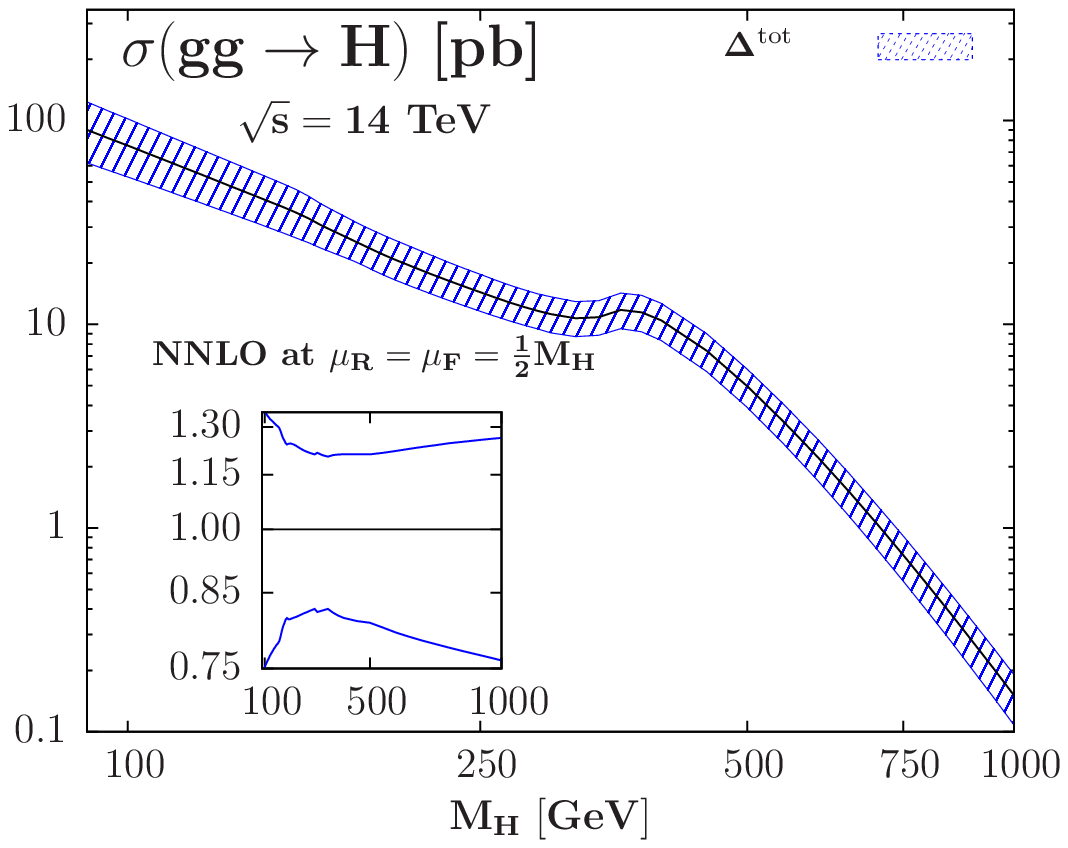}
\label{fig5a}}
\hspace{1cm}
\subfigure[]
{\includegraphics*[width=6cm,height=4.5cm]{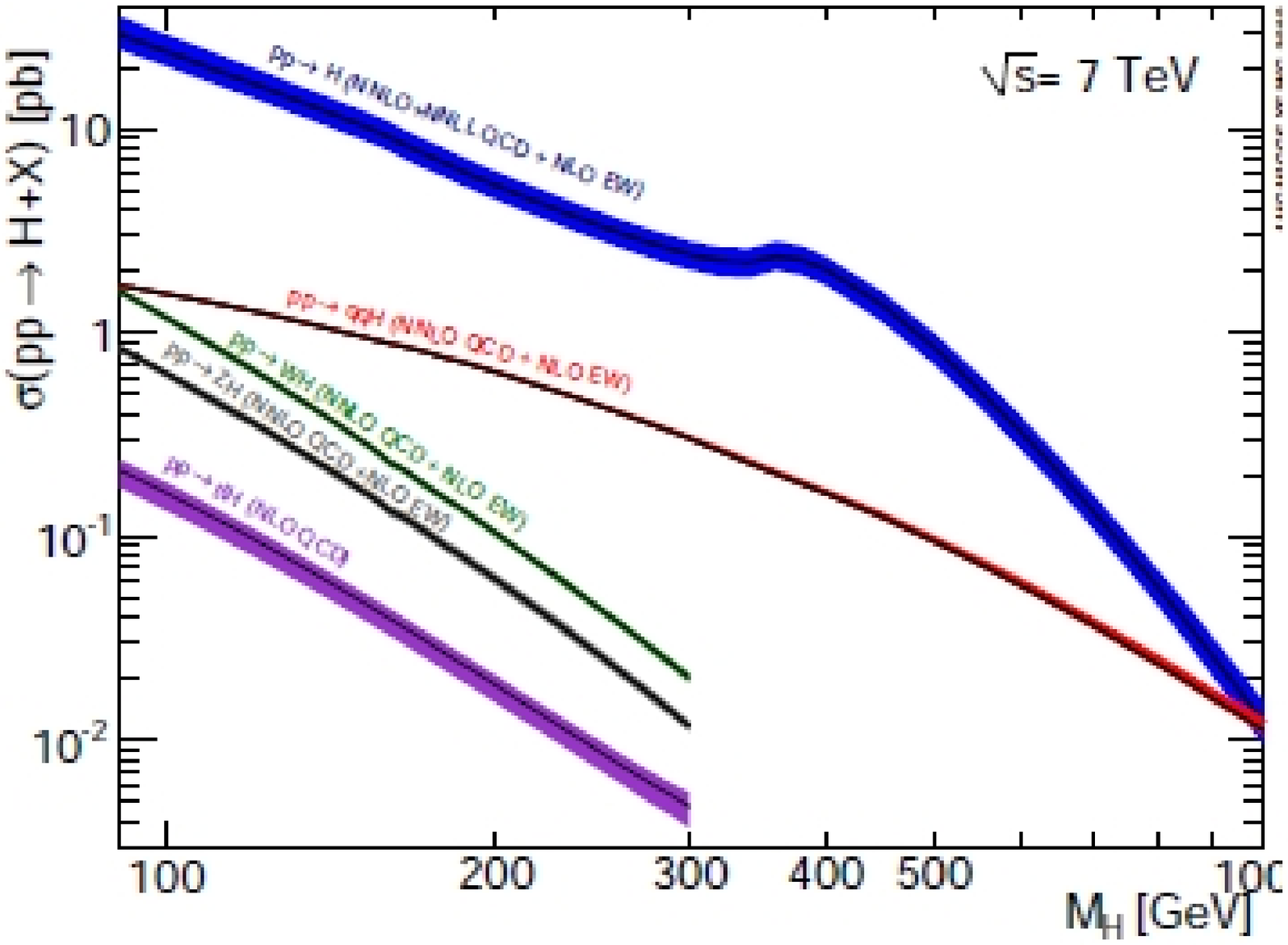}
\label{fig5b}}
\caption{Cross-sections for the $gg$ fusion process with all 
errors~\protect\cite{Baglio:2010ae}  for $\sqrt{s} = 14 $ TeV in the 
left panel and  cross-sections for all the relevant processes for
$\sqrt{s} = 7 $ TeV~\protect\cite{lhcxsection} in the right panel.}
\label{fig5}
\end{figure}

The most important mode of production at
the LHC is the $gg$ fusion, dominated by the top loop. The next to leading 
order (NLO) corrections have been computed both in the Effective Field 
Theory (EFT)  approach in the limit of infinite top mass and for finite 
heavy quark mass. Further, the next to next leading order (NNLO) corrections  
have been computed doing the three loop calculation. On top of it, the 
resummation of soft and collinear corrections has been performed at the 
next to next to leading logs (NNLL). The non factorisable EW and QCD 
corrections to the process  have also been computed and shown to be 
$\sim 5 \%$. The K-factor, defined as the ratio to the leading order (LO) 
crosssection,  for the dominant $gg$ fusion process for low Higgs masses, 
at the LHC, is 1.7 at NLO and grows to about 2 at NNLO, thus showing 
a good convergence of the perturbation series. The NNLO result has small 
dependence on the renormalisation and 
factorisation scale variations, the hallmark of stability of a perturbative 
QCD calculation. The cleanest prediction is for the $WH/ZH$ production, where
both the QCD and EW corrections have been computed and the resulting 
cross-section has a K-factor ~$\sim 1.2-1.3$ at NNLO.  The WW/ZZ fusion 
mechanism has the second largest cross-section at the LHC and would be very 
important for coupling/quantum number measurements once the Higgs boson has 
been discovered. In this case, the extraction of the signal for precision
measurements requires extensive cuts on the phase space and hence calculation
of higher order corrections to exclusive distributions is very important. Both
the QCD and EW corrections have been computed and the K-factors are found to
be modest. Equally important for the measurements of the couplings is the the
$t \bar t H$ production. Use of jet substructure method~\cite{tthjet} may yet 
revive the utility  of this channel.  The NLO corrections
to this $2 \rightarrow 3$ processes are now available and the scale variation
for the NLO result for $\sigma(pp \rightarrow t \bar t H)$  at the LHC is 
found to be rather modest ~$\sim 10$--$20 \%$.  
The two panels of Fig.~\ref{fig5} summarise the state of art 
predictions for the Higgs production cross section in $gg$ fusion for 
$\sqrt{s} = 14$ TeV from ~\cite{Baglio:2010ae} and $\sqrt{s} = 7$ TeV
for all the relevant production processes from ~\cite{lhcxsection}.

\begin{figure}[!hbtp]
\centering
\subfigure[]
{\includegraphics*[height=5.7cm,width=7.0cm]{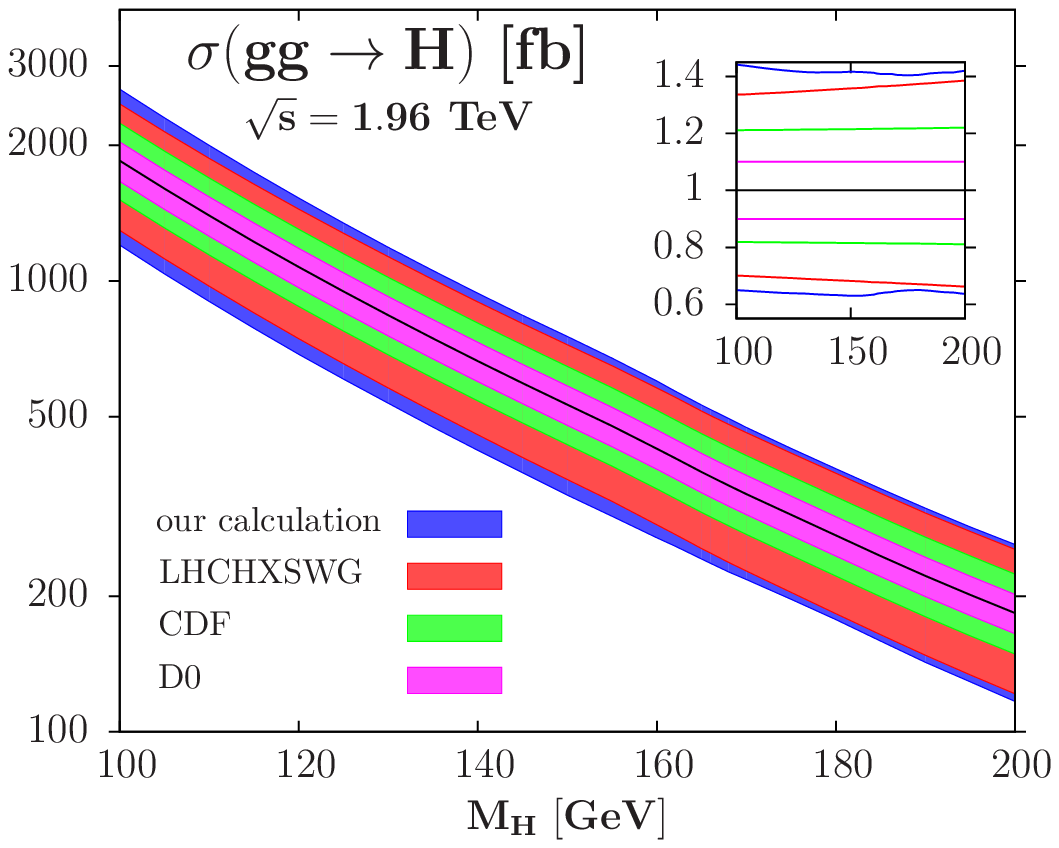}
\label{fig6a}}
\subfigure[]
{
\includegraphics*[height=6cm,width=8cm]{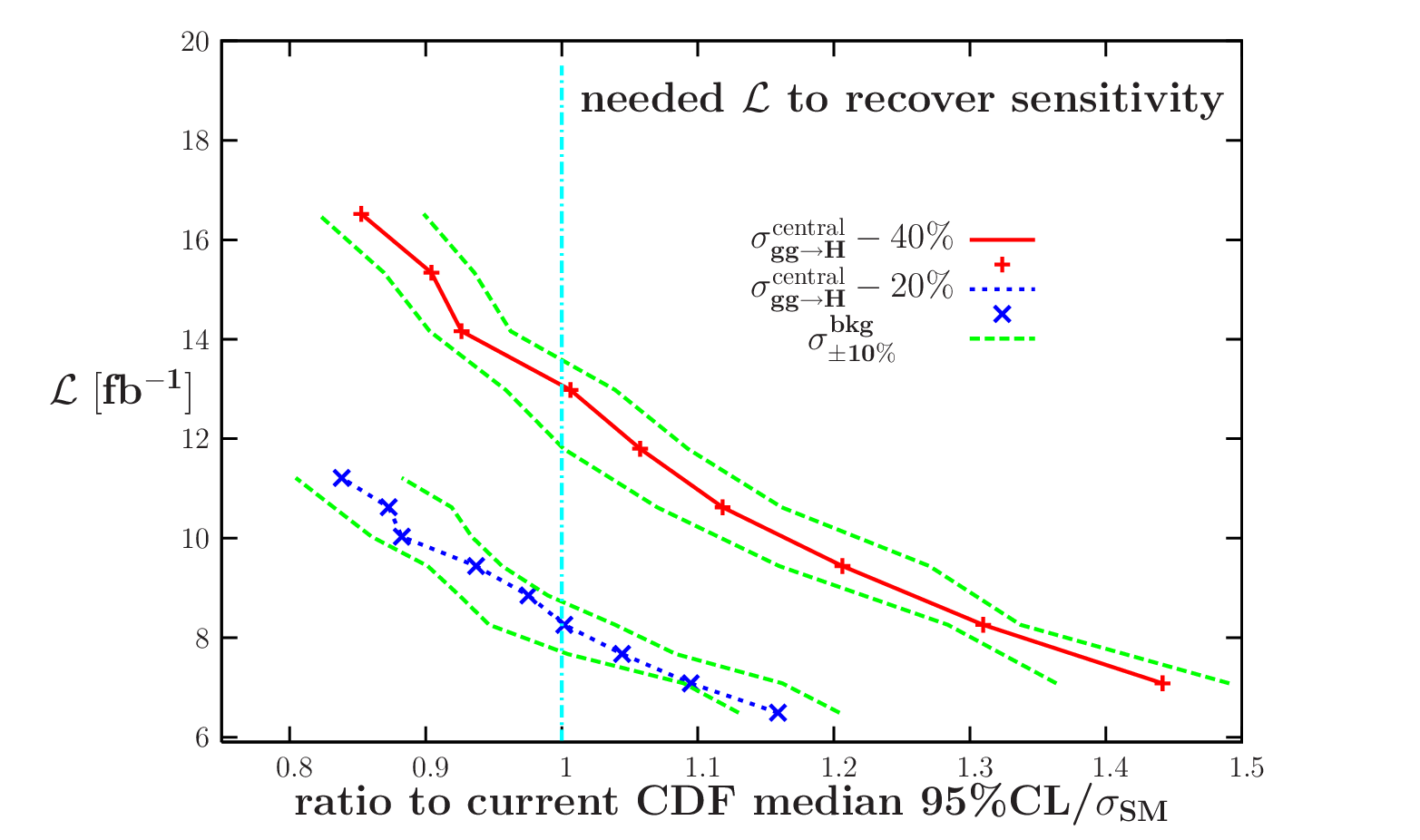}
\label{fig6b}}
\caption{The needed luminosity by the CDF experiment to recover the 
current sensitivity (with 5.9 fb$^{-1}$ data) when the $gg \rightarrow H 
\rightarrow  \ell \ell \nu \nu$ signal is lowered by $20\%$ and $40\%$ and with
a $\pm 10 \%$ change in the dominant $p \bar b \rightarrow W W$ background,
taken from ~\protect\cite{us}.}
\label{fig6}
\end{figure}

It is worth mentioning here that the situation about the theoretical 
uncertainties in the production cross-section of the $gg$ fusion process
at the Tevatron~\cite{Baglio:2010um,us} is quite different. In fact this 
process is observable at the Tevatron only because of the rather large NLO/NNLO
corrections it receives corresponding to a K-factor, defined as the ratio of 
the result of the higher order calculation to that of the LO, of 2(3)
at NLO (NNLO). This thus means that the range of variation of the common
factorisation and renormalisation scales in this case has to be somewhat larger
than for the LHC case, possibly  leading to a larger scale variation 
uncertainty in the cross-section. Further, the different parametrisations for 
the PDF's which correspond to different assumptions on these nonperturbative 
inputs, can  differ in the central value of the predicted 
cross-section~\cite{moch,us} by upto $40\%$  for Higss masses where the 
sensitivity is maximal.  The left panel in the Fig.~\ref{fig6}
taken from ~\cite{us} shows the Higgs production cross section 
$\sigma^{NNLO}_{gg \rightarrow H}$ at the Tevatron with the uncertainty band 
when all theoretical uncertainties are added as suggested in
Ref.~\cite{Baglio:2010um}. This uncertainty is compared the uncertainties 
quoted by the CDF and D0 experiments~\cite{tevlimits} as well as the 
uncertainty that results when the LHC procedure as 
suggested~\cite{lhcxsection} is adopted. In the insert, 
the relative size of the uncertainties compared to the central value are shown.
Thus we see that if one were to evaluate the theoretical uncertainties
for the Tevatron by the method prescribed in ~\cite{lhcxsection} one would get 
about $35 \%$ uncertainty in the cross-section as opposed to the 
$20\%$ and $10\%$ assumed in the CDF and D0 analyses
respectively~\cite{tevlimits}. These differences as well as the strong
dependence of the normalisation of the Higgs cross-section on the PDF,
suggest that one should critically evaluate the dependence of the exclusion 
bounds from the Tevatron shown in Fig.~\ref{fig3}, on the same.  In fact
analysis of Ref.~\cite{us} shows that if the true normalisation is indeed 
smaller by $\sim 40\%$ than that for the used MSTW parameterisation, then one 
might need upto a factor 2 higher luminosity to achieve the same exclusion. 
This situation is illustrated in the right hand panel (b) of this figure where 
the needed luminosity to recover the current sensitivity, if the cross-section 
should be lowered by $20 \%$ and $40 \%$, is indicated. This underlies the 
importance of having a complete assessment of the theoretical uncertainties as 
is presented in ~\cite{Baglio:2010um,Baglio:2010ae,lhcxsection,us}.

\section{LHC: projections and results}
\begin{figure}[!htbp]
\centering
\includegraphics*[width=7cm,height=6cm]{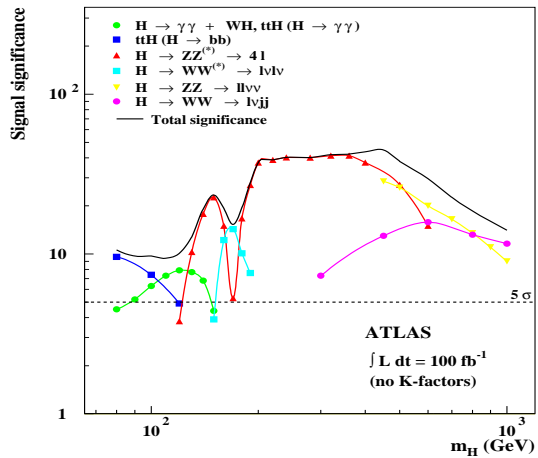}
\caption{The expected signal significance for different search channels
at the LHC with $14$ TeV, assuming no K-factors, for $100$ fb$^{-1}$ 
integrated luminosity. Taken from ATLAS TDR in \protect\cite{lhc}.}
\label{fig7}
\end{figure}

As said before at the LHC, $gg$ fusion is the dominant production
mechanism and the final state contributing to the discovery depends on 
the mass of the Higgs. Fig.~\ref{fig7} taken from the ATLAS TDR \cite{lhc},
shows the signal significance that one expected to achieve at the originally 
planned 14 TeV LHC,for an integrated luminosity of $100$ fb$^{-1}$, 
neglecting all the K-factors. This corresponds to the assertion that 
used to be made that a single experiment can discover the Higgs over the 
entire mass range allowed by theoretical considerations at $5 \sigma$ at the
14 TeV LHC.  

Now the LHC has been running at the lower energy of $7$ TeV, at 
a lower luminosity than planned but has already collected $35$ pb$^{-1}$
data per experiment thanks to the very good performance of the LHC machine.
It will now continue to run at $7$ TeV till end 2012.

\begin{figure}
\subfigure[]
{\includegraphics*[width=8cm,height=6cm]{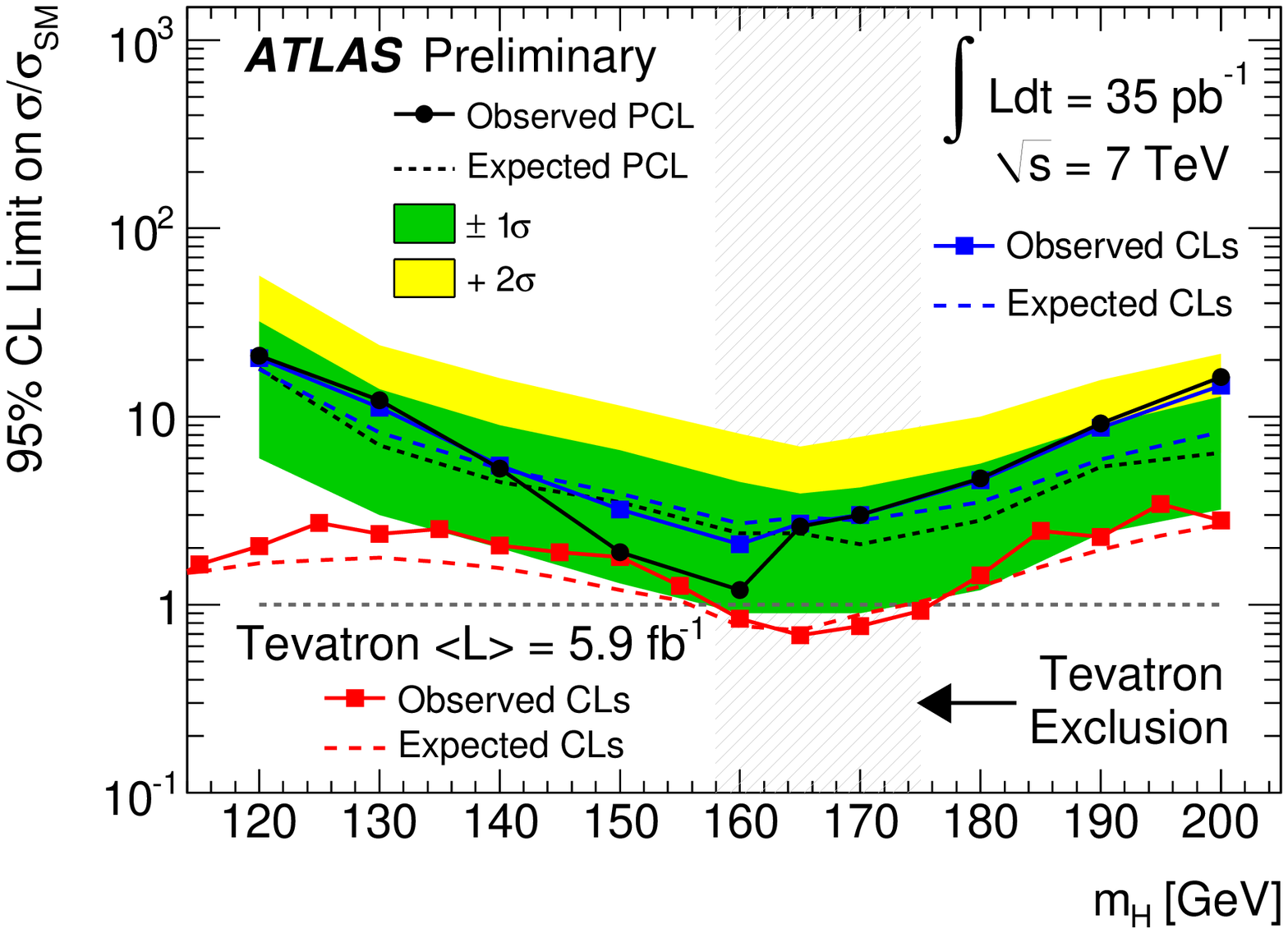}
\label{fig8a}}
\subfigure[]
{\includegraphics*[width=8cm,height=6cm]{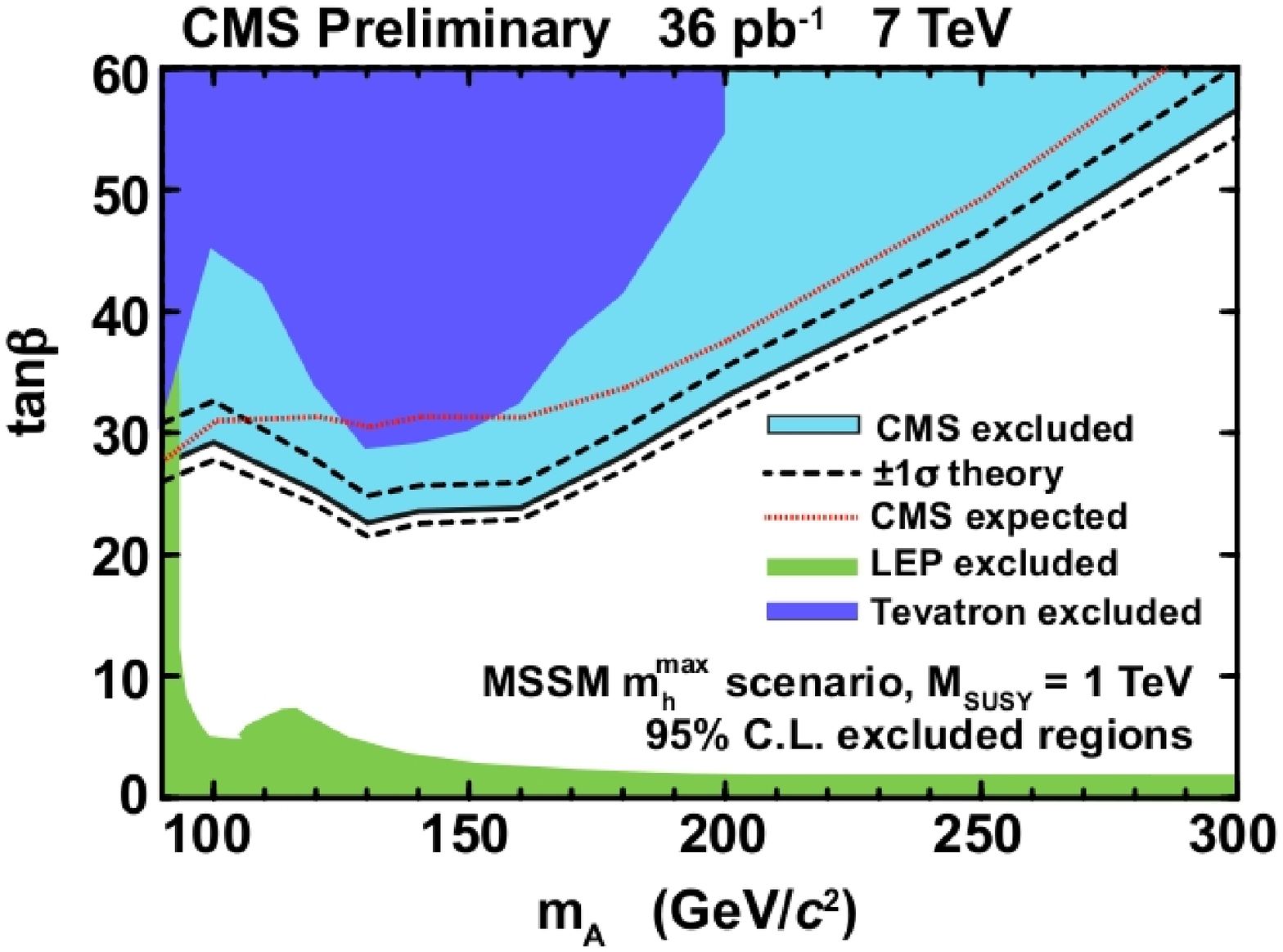}
\label{fig8b}}
\caption{Examples of the results for the SM Higgs and MSSM Higgs available
from the 35 pb$^{-1}$ luminosity at LHC taken from ~\protect\cite{atlasdata}
and ~\protect\cite{cmsdata} respectively.}
\label{fig8}
\end{figure}

Plots in Fig.~\ref{fig8} show that even with the very small amount of
data the LHC has started giving significant results. The left panel shows that 
the ATLAS collaboration is getting close to being sensitive to the SM Higgs
in the mass range around $160$ GeV  and has put limits on the 
cross-section ,at $95\%$ CL, of about  $1.2$ times the SM cross-section,
for Higgs mass around $160$ GeV.  Clearly, one has to watch this 
space now closely for future news.  

For the CMS results I have chosen the example of the SUSY 
Higgs about which I have not talked much in sections 2 and 3. Supersymmetry 
is one of the most popular and arguably the best motivated BSM physics
candidate. In the MSSM~\cite{susy-book} there exist 5 Higgs bosons, three 
neutrals and two charged, one of the three neutrals being a pseudoscalar.
An important difference from the SM case is that the lightest Higgs mass
is now constrained from above ($\sim 130$--$140$ GeV) and the Higgs mass 
bound depends on  some details of the specific  SUSY model and the 
parameters thereof. The heavier neutral Higgs bosons decay mostly into 
$b$ and $\tau's$ and thus the phenomenology is quite distinct. The production
cross-section for the inclusive production of the supersymmetric
Higgs in the process $ gg, b \bar b \rightarrow H$ with 
$H \rightarrow \tau \tau$, is considerably enhanced at large $\tan \beta$
cite{higgs-review} and is thus accessible even with low luminosity. 
The exclusion for the supersymmetric Higgs achieved by the CMS experiment
is shown in the figure in the right hand panel. The ATLAS exclusion
for the same seen in ~\cite{atlasdata} is similar. These results 
have already led to theoretical analyses of their implications not just for 
SUSY searches in general but also for the search of a light, SM
higgs at the LHC~\cite{Baglio:2011xz}.

\begin{figure}
\subfigure[]
{\includegraphics*[width=8cm,height=6cm]{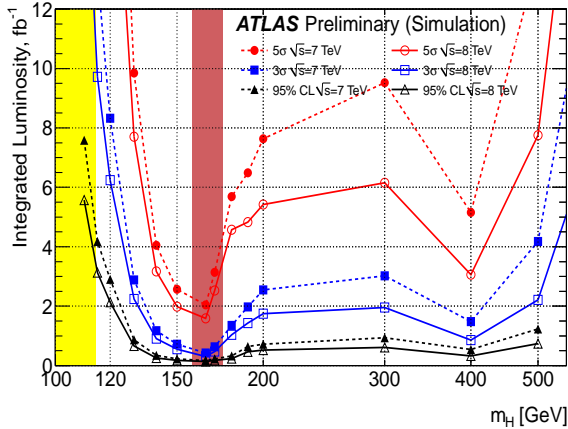}
\label{fig9a}}
\subfigure[]
{\includegraphics*[width=8cm,height=6cm]{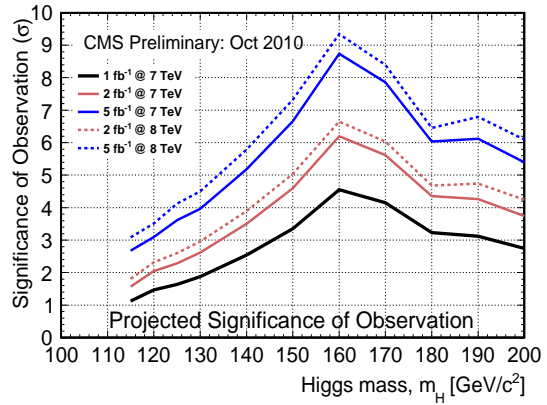}
\label{fig9b}}
\caption{
ATLAS simulation for the required integrated Luminosity for exclusion
at $95\%$ CL and discovery at 3 and 5 $\sigma$
level~\protect\cite{atlasproj} (left panel) and the  expected level of 
significance of observation at different integrated luminosities from CMS
simulation~\protect\cite{cmsproj}, as a function of $M_H$. Results are shown 
for both $\sqrt{s} = 7$ and $8$ TeV.}
\label{fig9}
\end{figure}

The LHC experiments seem to be performing amazingly well and the time gap 
between 
data taking and availability of results is indeed very short. It is therefore
important to know what are their projections now for the Higgs searches.  For 
more details I refer the readers to information available on the web pages
in Refs.~\cite{atlasproj} and \cite{cmsproj} respectively. The left panel of
Fig.~\ref{fig9} shows the luminosity required for $5$ ($3$) $\sigma$  Higgs
discovery and Higgs exclusion at $95 \%$  CL at centre of mass energy of 
7 and 8 TeV 
respectively, whereas the right panel shows the CMS version of the plot
of Fig.~\ref{fig7} but now for $\sqrt{s} = 7$ and $8$ TeV , for few selected
values of integrated luminosities. These figures show clearly that depending 
on the luminosity the LHC  machine manages to deliver, we would have
very significant information on the SM Higgs mass  by the end of the 2012 run.
This makes now for a very agonising wait indeed.

\section{Determination of Higgs properties and couplings}
As already stated, just discovering the Higgs at a particular mass 
and  the simultaneous results from the associated searches for BSM physics,
will begin to give indicative answers about whether the SM is a strongly 
coupled theory with a composite Higgs boson or a weakly coupled theory with
and elementary Higgs boson. But for a good scrutiny of the gauge group
representation to which the Higgs belongs and whether it is 'the' SM Higgs,
measurements of its couplings to the other SM particles, determination
of its spin and further determination of its CP property is quite essential. 
The standard wisdom~\cite{higgs-review} in this respect was that these are 
usually high luminosity measurements. For example, the studies of 
Ref.~\cite{zeppenfeld} had shown that with an integrated luminosity of
about $600~{\rm fb}^{-1}$, at $14$ TeV, it will be possible to measure various
couplings at  $\sim 20$--$30\%$ level for the SM Higgs. These results 
were confirmed  with a more sophisticated analysis recently~\cite{remy}. 

\begin{figure}[!htbp]
\subfigure[]
{\includegraphics*[width=8cm,height=6cm]{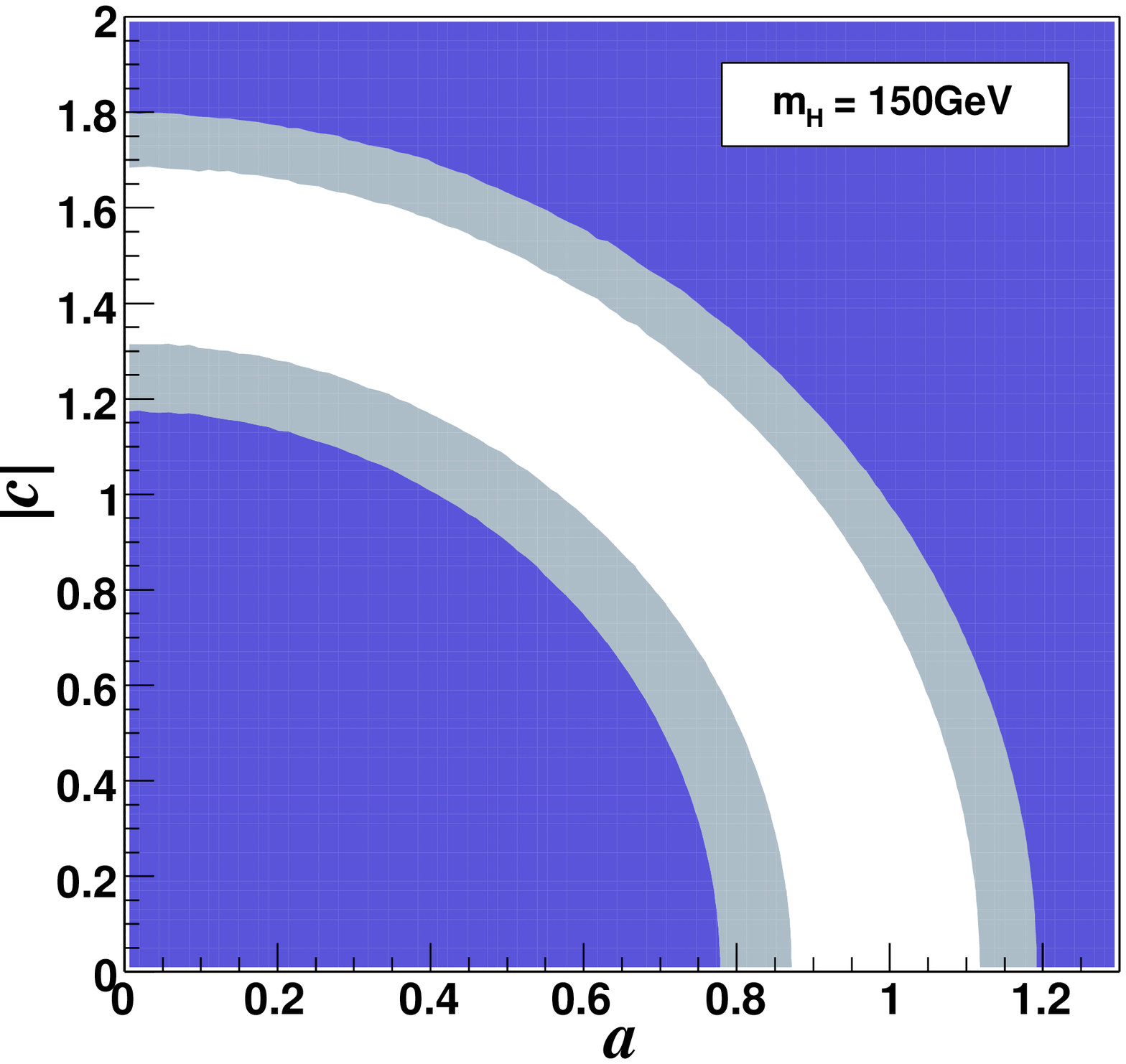}
\label{fig10a}}
\subfigure[]
{\includegraphics*[width=8cm,height=6cm]{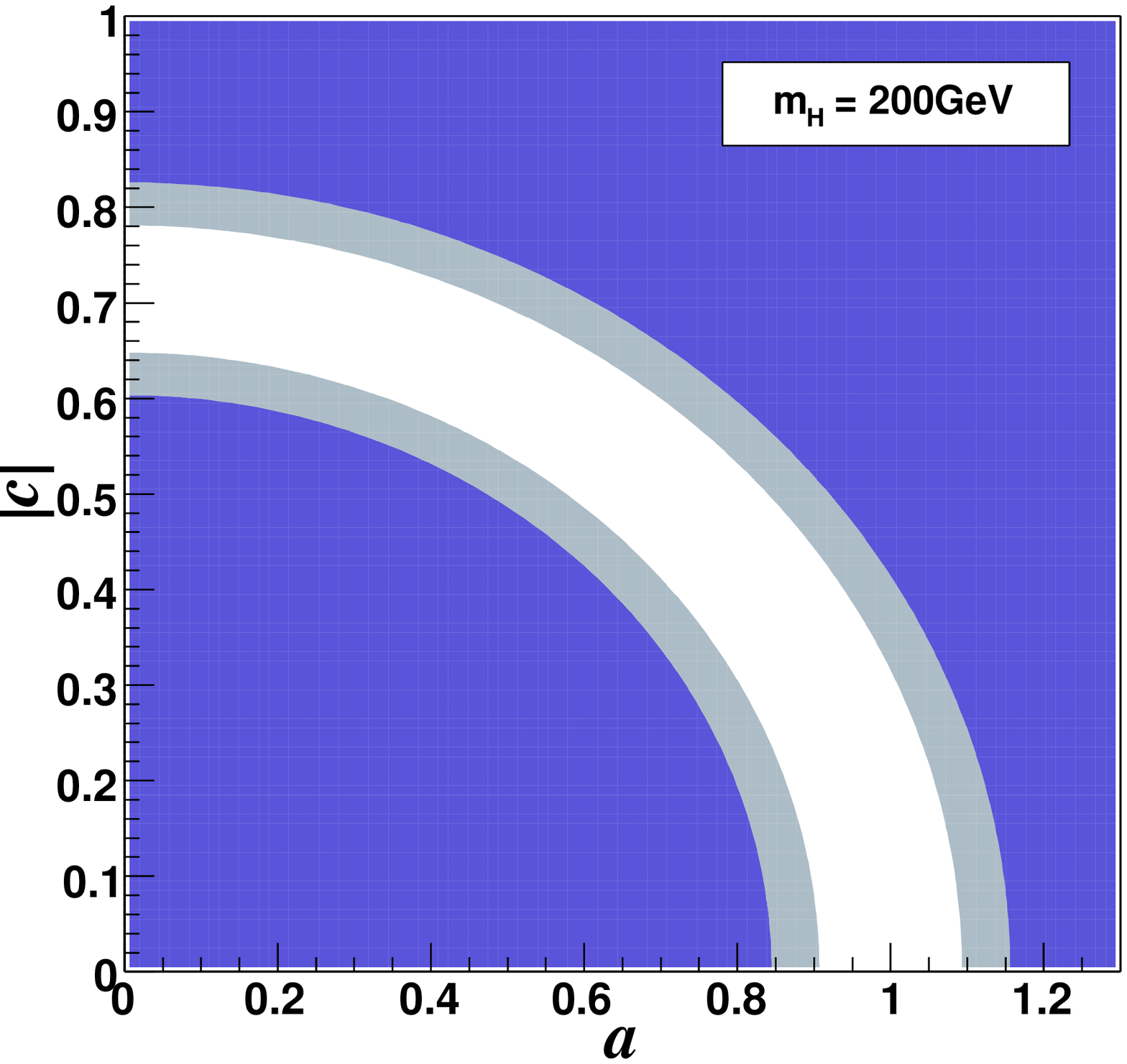}
\label{fig10b}}
\caption{The number of standard deviations from the SM which can be
obtained in the process $gg\to H\to Z^* Z^* \to 4$~{\it leptons}, as a
scan over the $(a,|c|)$ plane. The Higgs mass has been chosen to be
$150\,$GeV {\it (left)} and $200\,$GeV {\it (right)}. The white region
is where the deviation from the SM is less than $3 \, \sigma$; in the
light blue/light Grey region the deviation is between $3\,\sigma$ and
$5\,\sigma$; while for the dark blue/dark Grey region the deviation is
greater than $5\,\sigma$ for an integrated luminosity of $300$ fb$^{-1}$, taken
from \protect\cite{cpus}.}
\label{fig10}
\end{figure}

Another example is of
investigations of Ref.~\cite{cpus} which indicated that at $14$ TeV LHC, 
one would be able to establish some of the anomalous (CP violating) $HZZ$ 
couplings at $3$--$5~\sigma$ level, with $100$--$300$ fb$^{-1}$ integrated 
luminosity, if these couplings were of the same order of magnitude as the 
SM couplings. In Fig.~\ref{fig10} taken from Ref.~\cite{cpus}, are shown  
regions in the  $|c|$--$a$ coupling plane that can be probed by just a 
measurement of the width of the Higgs boson. Here the $HZZ$ vertex has 
been parametrized in the most general model independent way given by:
\begin{eqnarray}
 V_{HZZ}^{\mu \nu} \, =\,
\frac{ig m_Z}{\cos\theta_W} \left[ \,a\, g_{\mu\nu}
+  b \,\frac{p_\mu p_\nu}{m_Z^2}
+  c \,\epsilon_{\mu\nu\alpha\beta} \, \frac{ p^\alpha k^\beta}{m_Z^2}
\, \right],
\end{eqnarray}
in obvious notation for the different quantities appearing therein, $p,k$
representing the sum and difference of the four momenta of the two $Z$ bosons.

\begin{figure}
\centering
\includegraphics*[width=7cm,height=6cm]{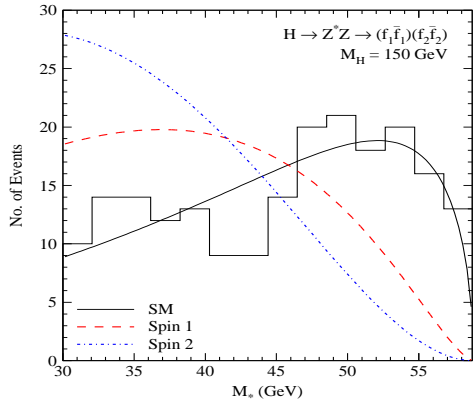}
\caption{Distribution in $M_{Z^*}$ for $H \rightarrow Z Z^*$
taken from \protect\cite{miller1}.}
\label{fig12}
\end{figure}

It was also demonstrated in ~\cite{miller1} that for a Higgs heavy enough to
have a reasonable branching ratio in the $Z Z^*$ channel, the shape of the
distribution in the invariant mass of the $\ell^+ \ell^-$ pair coming from
the $Z^*$ decay, can give clear information about the spin of the Higgs boson. 
The plot in Fig.~\ref{fig12} taken from Ref.~\cite{miller1}, shows the 
measurement possible for an integrated luminosity of $100$ fb$^{-1}$, 
at $\sqrt{s} = 14 $ TeV, the histogram showing the expected statistical error. 
It had also been shown that the distribution in the azimuthal angle between 
the planes of the two pairs of the decay leptons can also carry information 
about the spin and the parity of the decaying resonance~\cite{miller2}.
Recently there were investigations~\cite{derujula,gao}
which showed that more complicated, multivariate analyses might be able to do 
the job of establishing the $J^{PC}$ to be $0^{++}$ with  $\sim 3 \sigma$
significance  for $\lsim 10$ fb$^{-1}$ luminosity. 

Apart from the high luminosity, for the coupling measurements 
use of the $b \bar b $ final state is also essential 
and so is the possibility to  make a good 
measurement of the $t \bar t H$, $H \rightarrow b \bar b$  process.
The $t \bar t j j$ background seems to make the use of this channel
very difficult~\cite{lhc}. Hence any methods to improve the visibility
of this channel are welcome. As mentioned before, methods using substructure 
of jets have given new hope in both these issues~\cite{butterworth,tthjet}. 
I would therefore describe briefly this method.

\begin{figure}[!htbp]
\subfigure[]
{\includegraphics*[width=5cm,height=5cm]{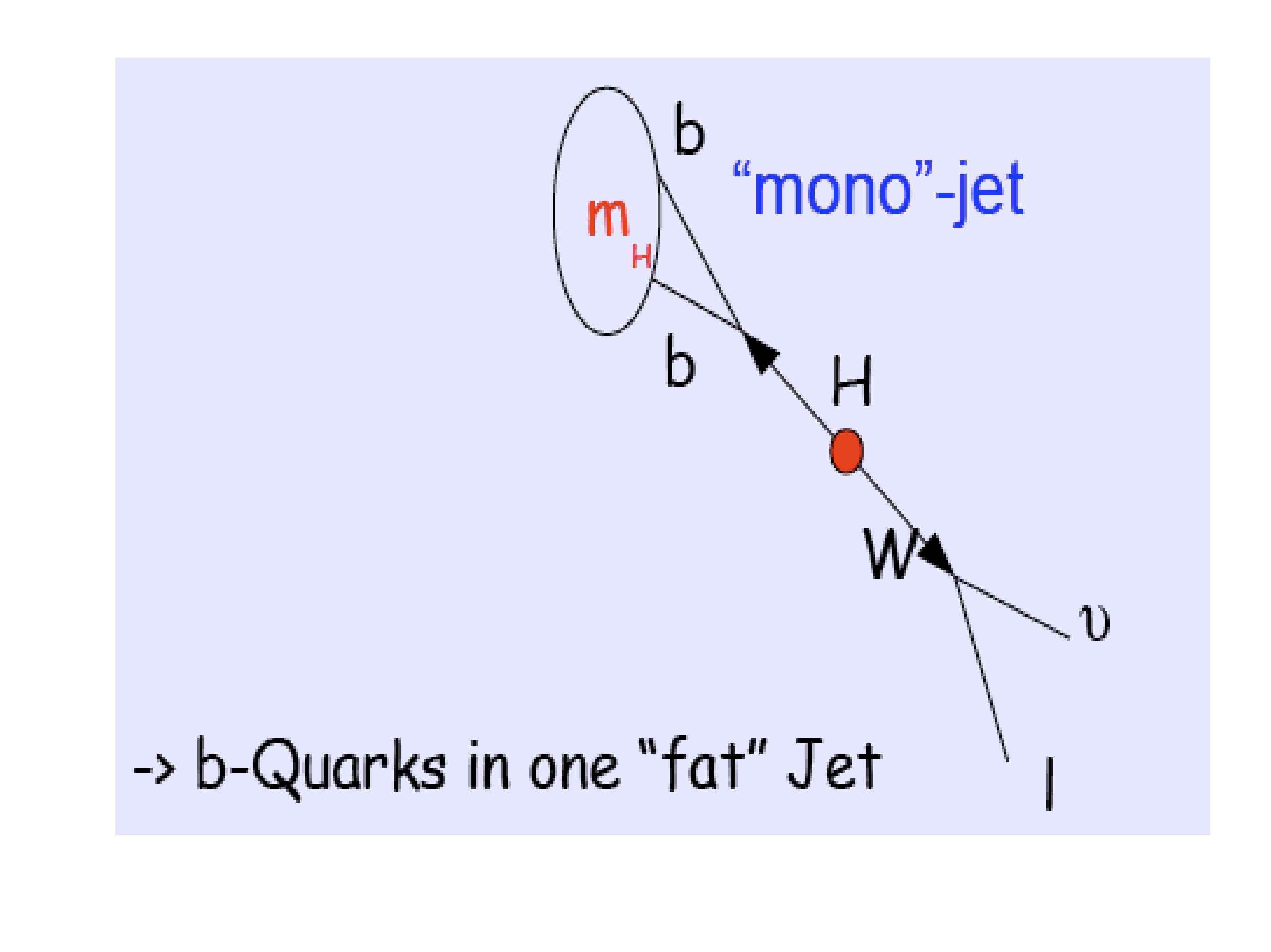}
\label{fig11a}}
\hspace{1cm}
\subfigure[]
{\includegraphics*[width=7cm,height=6cm]{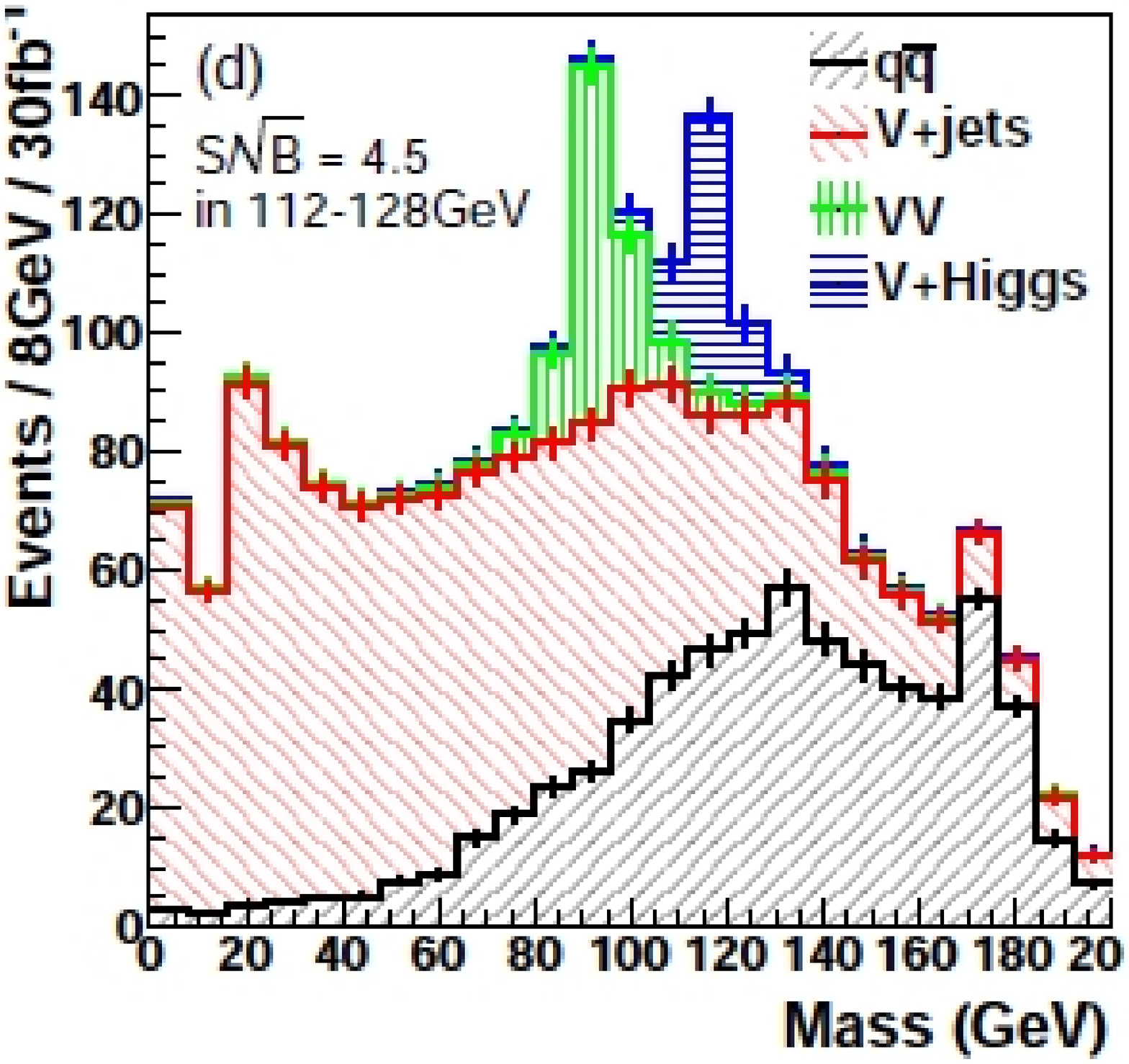}
\label{fig11b}}
\caption{Left panel shows a cartoon of a 'fat' jet from 
the $b \bar b$ decay of a large $P_T$ Higgs and the right panel shows
how the substructure analysis can help increase the $S/\sqrt{B}$ in the 
WH channel, taken from~\protect\cite{butterworth}.}
\label{fig11}
\end{figure}

The idea here is based on the fact that for high $p_T$ Higgs bosons, the 
$b \bar b$ decay products would emerge close to each other and hence will look 
in the detector to be a single, fat jet with large invariant mass. The jets
produced by QCD emission will not have this feature. Thus if one can
develop an algorithm to see if a fat, heavy jet is made of two fast objects
emitted close to each other, one can then reduce the QCD backgrounds to a low
level. For production processes like $WH,ZH$ where one has to select 
large $p_T$ Higgs bosons to get rid of the irreducible SM background anyway,
this technique seems to work quite nicely. The left panel in 
Fig.~\ref{fig11} shows a cartoon which illustrates this kinematical fact
and in the right panel is shown a plot from ~\cite{butterworth}, which indicates
the clean way in which the signal can be separated from the background for the
$WH$ case and $S /\sqrt{B}$ as high as 4.5 can be reached in this channel,
with $b \bar b$ final state, for $30$ fb$^{-1}$ of integrated luminosity for
a $120$ GeV Higgs boson.  These kinds of studies would be the future of Higgs
physics at the LHC once it has been discovered  through any one channel.

\section{Conclusion}
Thus we are now at a very exciting stage where in the next two years we should
expect either a $3 \sigma$ signal or a $95 \%$ exclusion over a very large 
range of the Higgs masses at the LHC. As said already the mass of the Higgs 
boson alone can give completely non trivial indications of the presence
or absence of BSM physics. Should the Higgs mass give an indication of 
the BSM physics,  in most cases the corresponding BSM physics should also
reveal itself in the  direct searches which would be taking place
alongside, for the value of the integrated 
luminosity we expect to have. Hence, the next two years of the Higgs physics
at the LHC should be very exciting indeed.  Precision measurements of the 
couplings, spin, parity, CP characteristic and determination of triple
Higgs couplings all  has to however, wait for higher luminosities,
higher energies and perhaps even for a leptonic collider~\cite{ilc}.

\section{Acknowledgments}
I wish to acknowledge the Department of Science and Technology of India,
for financial support  under the J.C. Bose Fellowship scheme under grant no. 
SR/S2/JCB-64/2007.  I would like to thank Abdelhak Djouadi not only for an 
enjoyable collaboration but also for a careful reading of the manuscript and 
useful comments.
%%%%%%%%%%%%%%%%%%%%%%%%%%%%%%%%%%%%%%%%%%%%%%%%%%%%%

\end{document}